\documentclass[
]{ceurart}

\usepackage{listings}
\begin{document}

\copyrightyear{2026}
\copyrightclause{Copyright for this paper by its authors.
  Use permitted under Creative Commons License Attribution 4.0
  International (CC BY 4.0).}

\conference{ALIT4ALL: 2nd International Workshop on AI Literacy Education For All, 
  July 2026, Seoul, Republic of Korea}

\title{Bridging Technical AI, Societal Impacts, and Workforce Competencies in AI Education}

\author[1]{Narges Zare}[%
email=nzare@charlotte.edu,
]
\cormark[1]

\author[1]{Divya Ramesh}[%
email=dramesh4@charlotte.edu,
]

\author[1]{Cori Faklaris}[%
email=cfaklari@charlotte.edu,
]

\address[1]{University of North Carolina at Charlotte, Charlotte, NC, USA}

\cortext[1]{Corresponding author.}
\begin{abstract}
As AI becomes embedded across everyday life and work, educators must help students connect technical knowledge with societal consequences and workplace responsibilities. Yet AI education often remains fragmented, with technical concepts, ethics, human-centered design, and workforce preparation taught separately. This work-in-progress presents a curriculum mapping framework that links technical systems, societal harms, and workforce competencies, beginning with institutional courses and expanding through an external 335-course registry. Early findings from six courses suggest that technical coverage is often strong, societal harms are present but unevenly integrated, and workforce competencies are rarely explicitly assessed. These findings speak to AI literacy efforts by showing that literacy cannot stop at awareness of AI concepts or harms; it must also include competencies for responsible action in AI-enabled contexts.
\end{abstract}

\begin{keywords}
AI Education \sep
AI Literacy \sep
Workforce Competencies \sep
Curriculum Mapping \sep
Human-Centered AI \sep
Societal Impacts of AI 
\end{keywords}

\maketitle

\section{Introduction}

Artificial intelligence (AI) systems increasingly shape decisions in hiring, lending, healthcare, education, policing, and online communication. When these systems fail, the consequences are both technical and social. Prior work has shown how algorithmic systems can reinforce discrimination \cite{Oneil2016, noble2018algorithms}, reproduce structural inequality \cite{eubanks2018automating}, and introduce opacity and governance risks \cite{jobin2019global}. As AI becomes embedded in public and private infrastructures, calls for responsible, human-centered approaches have intensified \cite{shneiderman2020hcai, yang2018mapping}.

Universities have introduced AI ethics modules and interdisciplinary coursework, while professional organizations increasingly emphasize ethics, fairness, and workforce preparation in computing curricula \cite{ACM_IEEE_AAAI_CS2023_Beta, cra2024empowering}. Yet AI education often remains fragmented, with technical concepts taught separately from ethics, policy, human-centered design, and professional skill development \cite{10.1145/3328778.3366825, 10.1145/3341164}. Technical units frequently emphasize models and optimization, while discussions of bias, governance, and societal impact remain abstracted from implementation constraints.

This fragmentation reflects a broader sociotechnical challenge. Technical abstraction can obscure how design decisions translate into social consequences \cite{10.1145/3287560.3287598}, while fairness cannot be addressed through technical metrics alone \cite{green2018myth}. Despite growing attention to fairness, accountability, and transparency, there is limited empirical work examining whether technical instruction, societal harms, and workforce competencies are structurally aligned within AI-related curricula.

This work-in-progress proposes an integrated perspective that links (1) technical systems, (2) societal harms, and (3) workforce competencies. While this paper speaks to AI literacy education, our analysis focuses more broadly on AI-related curricula, including technical, human-centered, ethics-focused, and policy-oriented courses. We use this broader scope because societal impacts and workforce competencies are often distributed across multiple course types rather than contained within a single AI literacy course. Through cross-course curriculum mapping, we trace how technical mechanisms connect to lived consequences and professional skills, surfacing patterns and misalignments that can inform curriculum design.

Our research question is:

\begin{quote}

\textbf{How can AI-related curricula more systematically align technical systems, societal harms, and workforce competencies?}

\end{quote}

\subsection{Contributions}
This paper contributes: (1) an initial cross-course curriculum mapping approach linking technical systems, societal harms, and workforce competencies; (2) an emerging alignment model for case-based curriculum mapping; (3) preliminary findings showing strong technical coverage, uneven integration of societal harms, limited explicit assessment of workforce competencies, and a mismatch between technical rigor and social critique; and (4) early design implications for AI education.







\section{Background and Motivation}

AI literacy is one important part of AI education, but broader AI curricula must also prepare students to connect technical systems with societal impacts and workplace responsibilities. Prior work shows persistent challenges in integrating ethics into technical curricula, where ethics is often treated as a standalone topic rather than embedded within core instruction \cite{10.1145/3328778.3366825, 10.1145/3330794, 10.1145/3341164}. Sociotechnical research further shows that abstraction can obscure how design choices produce downstream harms \cite{10.1145/3287560.3287598}, and that technical fairness solutions alone are insufficient without contextual understanding \cite{green2018myth}. Human-centered AI emphasizes integrating technical and human considerations such as collaboration, interpretability, and user-centered design \cite{10.1145/3290605.3300233, yang2018mapping, shneiderman2020hcai}. At the same time, workforce-oriented research highlights the importance of communication, interdisciplinary collaboration, and responsible innovation \cite{cra2024empowering}. We use workforce competencies to refer to transferable professional skills needed to work responsibly with AI systems, including communication, problem framing, critical evaluation, interdisciplinary collaboration, stakeholder reasoning, and responsible innovation. Together, these gaps motivate an integrated approach to AI education that connects technical systems, societal harms, and workforce competencies through curriculum mapping and case-based instructional design.

\section{Methods}

\subsection{Study Design}

This work-in-progress uses a qualitative curriculum mapping approach to examine how AI-related curricula align technical systems, societal harms, and workforce competencies. The goal is to identify early patterns of alignment and misalignment across courses rather than provide exhaustive coverage.

\subsection{Course Selection and Dataset Construction}
The dataset was constructed in two stages. We began with three coordinated AI-related courses within an AI degree program: AI Literacy, Human-Centered AI, and Social Technology Design. To reduce single-institution bias, we expanded the dataset using publicly available AI and technology-ethics curriculum registries, including Casey Fiesler’s public technology ethics syllabus collection as a source for identifying external courses, not as the analytical framework for this study \cite{fiesler2018techethics}. We screened course titles using AI-related terms such as AI, machine learning, algorithmic, generative AI, LLM, computer vision, NLP, recommender systems, human-centered AI, human-AI, and sociotechnical.

Candidate courses were then manually reviewed using syllabi, course descriptions, weekly topics, learning objectives, and assignments. Courses were included if they centered AI systems, engaged societal impacts such as bias, fairness, privacy, or governance, and provided enough detail for structured analysis. This process produced 16 confirmed AI-relevant courses from institutional and external sources. The present paper reports early findings from six courses mapped in depth.

We then applied keyword screening and manual validation. Courses were included if they treated AI systems as a primary instructional focus, engaged societal impacts such as bias, fairness, privacy, or governance, and provided sufficient detail for structured analysis. Courses focused solely on technical implementation or ethics without AI system engagement were excluded. The confirmed dataset includes 16 courses; for courses without public syllabi, we contacted instructors to request materials, and these remain pending validation.

We analyzed syllabi, weekly topics, learning objectives, assignments, and course descriptions. The dataset spans multiple disciplines and course levels. We also used curated sociotechnical case studies and external competency frameworks to support cross-course comparison.

\subsection{Mapping and Analysis}

We operationalized a three-part alignment model: (1) technical systems, (2) societal harms, and (3) workforce competencies. Course materials were iteratively coded across these dimensions. Workforce competencies were identified when syllabi, learning objectives, or assignments explicitly named these skills, or when course activities required students to practice them through projects, presentations, stakeholder analysis, critique, or design work. We then conducted cross-course comparison to identify patterns of co-occurrence and gaps, focusing on whether societal harms were linked to specific technical mechanisms and whether competencies were explicitly assessed.

\subsection{Scope and Limitations}

As a work-in-progress, this analysis draws from an evolving 16-course dataset. The present study focuses on an initial subset examined in depth to surface early patterns rather than definitive claims. Our unit of analysis is both the individual course and the broader curriculum landscape. We do not assume that every course should address technical systems, societal harms, and workforce competencies equally; instead, we examine how these dimensions are connected within and across courses.

\section{Early Findings}

\begin{figure}[t]

  \centering

  \includegraphics[width=\linewidth]{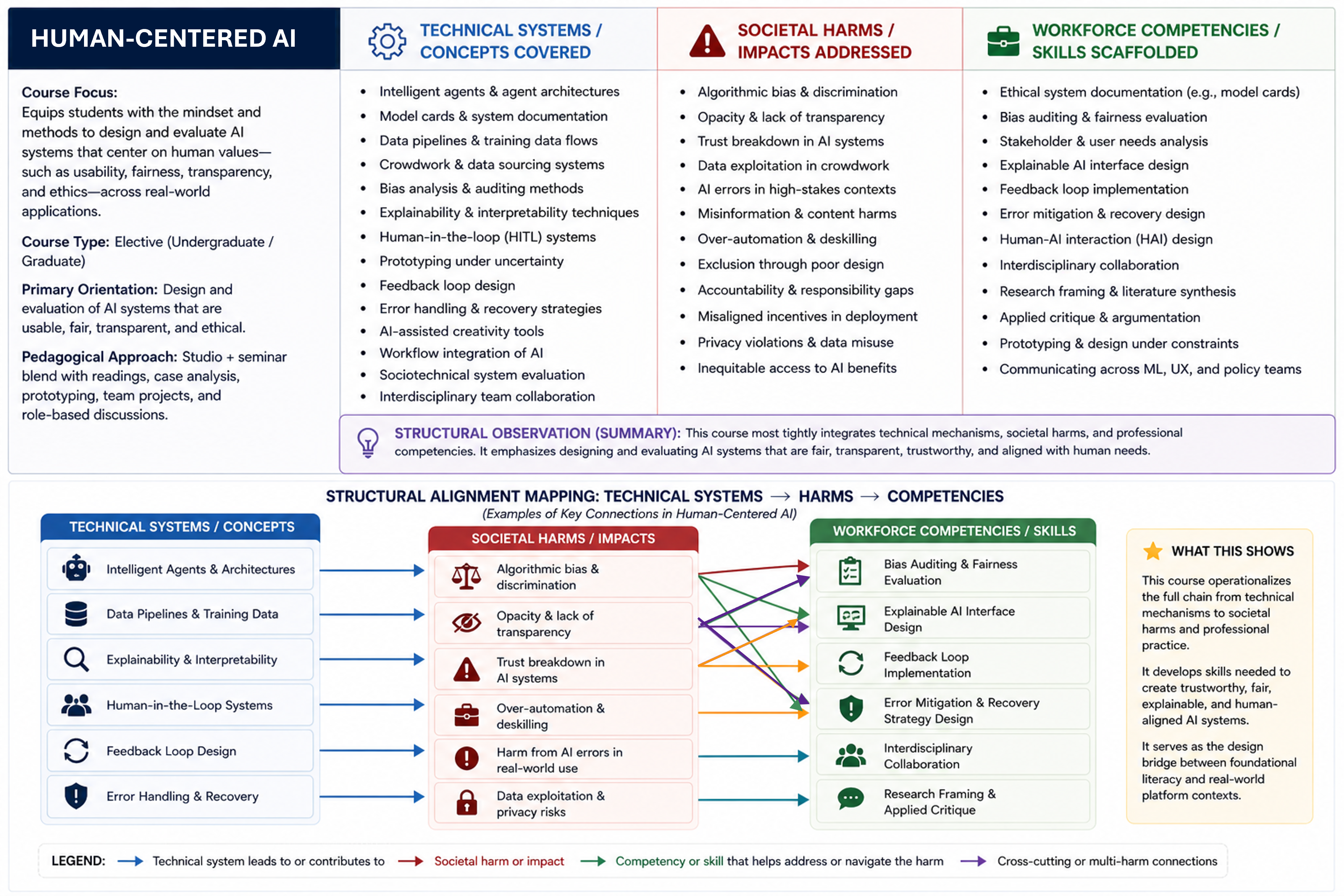}

  \caption{Example structural alignment mapping showing how technical systems connect to societal harms and how competencies are scaffolded to address those harms within a Human-Centered AI course.}

  \label{fig:alignment_example}

\end{figure}

We report early findings from six courses mapped in depth: AI Literacy for Professional Success, Human-Centered AI, Social Technology Design, Fairness in Machine Learning, Responsible AI, Law, Ethics \& Society, and Computational Ethics for NLP. Across these courses, we observed several recurring patterns. Figure~\ref{fig:alignment_example} shows an example mapping from the Human-Centered AI course linking technical systems, societal harms, and workforce competencies.

\subsection{Technical Systems Are Often Well Represented, But at Different Levels of Depth}

Technical coverage was generally strong, though depth varied across courses. Courses such as Fairness in Machine Learning and Computational Ethics for NLP connected technical components like supervised learning, embeddings, bias detection, and NLP pipelines to harms such as misinformation and profiling. In contrast, AI Literacy, Social Technology Design, and Responsible AI courses more often discussed AI systems at a broader level, including AI tools, recommender systems, platform governance, content moderation, and AI lifecycle management. This variation is not necessarily a weakness; it reflects differences in course purpose, audience, and disciplinary orientation. Analytically, the important issue is whether each course’s level of technical detail is connected to societal harms and professional competencies in ways that support responsible AI practice.

\subsection{Societal Harms Are Present, But Unevenly Integrated}

All six mapped courses addressed societal harms such as bias, privacy risks, misinformation, surveillance, labor displacement, and accountability gaps. However, these harms were integrated differently across courses. Some courses, such as Human-Centered AI and Computational Ethics for NLP, connected harms directly to design, data, model, or evaluation choices. Others, such as Social Technology Design, emphasized platform-level critique: discussions of harms associated with large-scale sociotechnical systems, such as content moderation, recommender systems, data governance, surveillance, or platform accountability, without always tracing those harms back to specific data, model, design, or evaluation choices. We do not argue that every harm must be reduced to an algorithmic mechanism; rather, structural connection means making clear how technical, organizational, and social decisions interact to produce or mitigate harm.

\subsection{Workforce Competencies Are the Least Explicitly Assessed}

Workforce competencies appeared across courses, including communication, critical evaluation, interdisciplinary collaboration, ethical reasoning, stakeholder reasoning, and responsible innovation. However, these competencies were often implied through projects, discussions, or presentations rather than explicitly named or assessed. Technical courses tended to emphasize analytical competencies such as model evaluation, fairness metrics, and error analysis, while human-centered and policy-oriented courses emphasized communication, governance, stakeholder reasoning, and interdisciplinary collaboration. Across the six courses, the clearest gap was not whether competencies appeared, but whether they were explicitly assessed and connected to technical learning outcomes.

\subsection{Misalignment Appears Between Technical Rigor and Social Critique}

A recurring pattern across the mapping was a mismatch between technical rigor and social critique. While some courses (e.g., Figure~\ref{fig:alignment_example}) showed strong alignment across technical systems, societal harms, and workforce competencies, this pattern was inconsistent across the curriculum. Courses with strong algorithmic depth often provided clearer explanations of harms but gave less attention to communication, implementation contexts, or governance practice. In contrast, courses with stronger societal critique often discussed harms in depth without fully explaining the technical mechanisms behind them. The strongest alignment appeared in case-based or project-based courses that connected systems, harms, and competencies within the same assignment structure. Together, these findings suggest that AI curricula require more deliberate connections across these dimensions, whether within individual courses or across coordinated course sequences.

\section{Discussion}

These early findings highlight a persistent gap between how AI is taught and how it is used in practice. Prior work has shown that separating ethics from technical instruction limits students’ ability to connect design decisions to real-world outcomes \cite{10.1145/3328778.3366825, 10.1145/3330794}. Our results illustrate this gap: even when societal harms are discussed, they are not always tied back to specific technical mechanisms.

One key implication is that simply adding ethics content is not enough. As prior research suggests, technical abstraction can obscure how systems produce harm \cite{10.1145/3287560.3287598}, and fairness cannot be addressed through metrics alone \cite{green2018myth}. Our findings reinforce this by showing that courses often either focus on technical rigor or social critique, but rarely integrate both in a structured way. This suggests that AI education should move beyond topic coverage toward explicit alignment between systems, impacts, and decision-making skills.

Another important insight relates to workforce preparation. Although competencies such as communication, collaboration, and ethical reasoning are widely recognized as essential \cite{cra2024empowering}, they are often not explicitly assessed in AI courses. This creates a gap between what students learn and what they are expected to do in real-world settings. Our results suggest that competencies should be embedded directly into technical assignments, rather than treated as separate or secondary outcomes.

These findings highlight the value of case-based, cross-course approaches. When courses use shared cases or projects to connect systems, harms, and skills, alignment becomes more visible and actionable. This aligns with human-centered AI perspectives that emphasize integrating technical and human considerations in system design \cite{10.1145/3290605.3300233, shneiderman2020hcai}.

\section{Conclusion and Next Steps}

This study focuses on a subset of courses and publicly available materials, which may not fully reflect classroom practice. Next, we will expand the dataset by incorporating additional curriculum repositories and teaching resources, complete pending syllabus collection, refine the coding scheme, and develop clearer measures of alignment across institutions. Overall, this work suggests that AI literacy efforts should connect to broader AI education by linking technical systems, societal harms, and workforce competencies across individual courses and coordinated course sequences.

\section*{Declaration on Generative AI}
The authors used ChatGPT and Gemini for brainstorming, drafting, and language refinement. The authors reviewed all AI-assisted content and take responsibility for the final paper.

\bibliographystyle{ceurmal}
\bibliography{main}
\end{document}